\documentclass{aa}
\usepackage[dvips]{graphicx}

\begin{document}

   \thesaurus{06         
              (03.13.1;  
               09.04.1;  
               09.08.1;  
               11.05.2;  
               11.09.4;  
               13.09.3)} 
   \title{Emission from dust in galaxies: metallicity dependence}

   \author{H. Hirashita
           \inst{1}\fnmsep\thanks{Research Fellow of the Japan Society
                               for the Promotion of Science},
           A. K. Inoue\inst{1}, H. Kamaya\inst{1},
          \and
          H. Shibai\inst{2}
          }

   \offprints{H. Hirashita}

   \institute{Department of Astronomy, Faculty of Science, Kyoto
              University, Sakyo-ku, Kyoto 606-8502, Japan\\
              (hirasita,inoue,kamaya@kusastro.kyoto-u.ac.jp)             
         \and
              Division of Particle and Astrophysical Science,
              Graduate School of Science, Nagoya
              University, Chikusa-ku, Nagoya 464-8602, Japan
              (shibai@phys.nagoya-u.ac.jp)
             }

   \date{Received 26 August 2000 / Accepted 6 November 2000}

   \authorrunning{Hirashita et al.}
   \maketitle

   \begin{abstract}
Infrared (IR) dust emission from galaxies is frequently used as an
indicator of star formation rate (SFR). However, the effect of the
dust-to-gas ratio (i.e., amount of the dust) on the conversion law
from IR luminosity
to SFR has not so far been considered. Then, in this paper, we 
present a convenient analytical
formula including this effect.
In order to obtain the dependence on the dust-to-gas ratio,
we extend the formula derived in our previous paper,
in which a theoretical formula converting IR luminosity
to SFR was derived. That formula was expressed as
${\rm SFR}/(M_\odot~{\rm yr}^{-1})=\{ 3.3\times 10^{-10}(1-
\eta )/(0.4-0.2f+0.6\epsilon )\} (L_{\rm IR}/L_\odot )$, where
$f$ is the fraction of ionizing photons absorbed by hydrogen,
$\epsilon$ is the efficiency of dust absorption for nonionizing
photons, $\eta$ is the cirrus fraction of observed dust
luminosity, and $L_{\rm IR}$ is the observed luminosity of
dust emission in the 8--1000-$\mu$m range. 
Our formula explains the IR excess 
of the Galaxy and the Large Magellanic Cloud.
In the current paper, especially,
we present the metallicity dependence of our conversion law
between SFR and $L_{\rm IR}$.
This is possible since both $f$ and $\epsilon$ can be estimated 
via the dust-to-gas ratio, which is related to metallicity.
We have confirmed that
the relation between the metallicity and the
dust-to-gas ratio is applied to both giant and dwarf galaxies.
Finally, we apply the result
to the cosmic star formation history. We find that the comoving
SFR at $z\sim 3$
calculated from previous empirical formulae is underestimated
by a factor of 4--5.

   \end{abstract}

%

\section{Introduction}

When we want to know the entire evolutionary history of a galaxy,
its color and metallicity are important diagnostic
quantities (e.g., Tinsley \cite{tinsley72}).
The evolution of the color and the metallicity results from 
the superposition of successive star formation 
during the lifetime of the galaxy.
Thus, if we wish to reveal the evolution of galaxies,
we always need to estimate the star formation rate (SFR) on a
galaxy-wide scale. There are many methods to estimate the SFR
of galaxies from observational quantities (Kennicutt
\cite{kennicutt98a}). In this paper,  
we are  especially interested in the conversion formula from
infrared (IR) luminosity of the galaxies to their present SFR.
The IR luminosity originates from dust that is heated by stellar
radiation field. Kennicutt (\cite{kennicutt98b})
estimated SFR from IR luminosity for starburst galaxies.
Buat \& Xu (\cite{buat96}) adopted an empirical approach by utilizing the
observed relation between the ultraviolet (UV) luminosity
and the IR luminosity. 

Recently, Inoue et al.\ (\cite{inoue00a}, hereafter IHK00) have derived a
theoretical conversion formula from IR luminosity
to the SFR by developing a standard model of H {\sc ii} regions
by Petrosian et al.\ (\cite{petrosian72}).
The present paper examines how metallicity (or dust-to-gas ratio)
affects the conversion
factor of IHK00. It is possible because
the effect of the metallicity is well parameterized in their formula.
Here, we present it as
\begin{eqnarray}
\frac{{\rm SFR}}{M_\odot~{\rm yr}^{-1}}=
\frac{3.3\times 10^{-10}(1-\eta )}{0.4-0.2f+0.6\epsilon}\,
\frac{L_{\rm IR}}{L_\odot}\, ,\label{inoue}
\end{eqnarray}
where $f$ is the fraction of ionizing photons absorbed by
hydrogen, $\epsilon$ is the efficiency of dust absorption for
nonionizing photons, $L_{\rm IR}$ is the observed luminosity of
dust in the wavelength range of
8--1000 $\mu$m, and $\eta$ is the cirrus fraction of $L_{\rm IR}$.
In Eq.\  (\ref{inoue}), $f$ and $\epsilon$ depend on
the dust-to-gas ratio\footnote{In this paper, we focus our main
attention on only $f$ and $\epsilon$, although we should keep 
in mind that the ($1-\eta$) may depend on the dust-to-gas ratio.
The effect of ($1-\eta$) on SFR is discussed in Appendix A.}. 
According to Hirashita (1999a, b),
the dust-to-gas ratio is expressed as a function of metallicity (see
also Dwek \cite{dwek98}; Lisenfeld \& Ferrara \cite{lisenfeld98}).
Therefore, we find easily that the relation
between $L_{\rm IR}$ and SFR depends on metallicity via $f$ and
$\epsilon$.

In this paper, we quantitatively examine the importance
of the metallicity to our conversion law, the conclusion being
that the metallicity dependence is not always negligible when we
compare SFRs of
very young galaxies and those of present galaxies.
First, we examine the formula by
IHK00 in Sect.\ \ref{sec:sfr}, focusing on the effect
dust-to-gas ratio. Then, in Sect.\ \ref{sec:metal},
we present the metallicity
dependence of the IHK00's formula. In Sect.\ \ref{sec:cosmic}, we extend
our discussion to the cosmic star formation history. Finally,
we summarize our conclusions in Sect.\ \ref{sec:conclusion} .


\section{Dependence on dust-to-gas ratio}\label{sec:sfr}
In this section, we derive the metallicity dependence of the
conversion formula (Eq.\ (\ref{inoue})). It is convenient
to define the factor $C_{\rm IR}$ as
\begin{eqnarray}
{\rm SFR}   &    =   & C_{\rm IR}L_{\rm IR}\, ,\label{eq:conversion} \\
C_{\rm IR}  & \equiv &
    \frac{3.3\times 10^{-10}(1-\eta )}{0.4-0.2f+0.6\epsilon}~
    [M_\odot~{\rm yr}^{-1}~L_\odot^{-1}]\, .\label{eq:c_ir}
\end{eqnarray}
We focus on the dependence of $C_{\rm IR}$ on metallicity.
The dependence of $C_{\rm IR}$ on dust-to-gas ratio is
included through
$f$ and $\epsilon$ as will be described in Sects.\ \ref{subsec:f}
and \ref{subsec:eps}. 
We will check our formula observationally in Sect.\ \ref{subsec:check}.
Then, using the relation between the dust-to-gas
ratio and the metallicity, we will obtain the metallicity
dependence of $C_{\rm IR}$ in the next section. 
We assume Case B (Baker \& Menzel \cite{baker38}), where the optical
depths for the Lyman series are large enough.

\subsection{Dependence of $f$ on dust-to-gas ratio}
\label{subsec:f}

The dependence of $f$ (Eq.\ (\ref{eq:c_ir})) on dust-to-gas ratio is
obtained from Spitzer (\cite{spitzer78}, hereafter S78). We define
$\tau_{\rm S,\, d}$ and
$\tau_{\rm S,\, H}$ as the optical depths of the dust and of
the neutral hydrogen atoms, respectively, for the Lyman-continuum
photons over a path length equal to the Str\"{o}mgren radius,
$r_{\rm S}$ (see S78 or Eq.\ (\ref{eq:stromgren}) for the definition
of $r_{\rm S}$).
Since the radius of a dust-free H {\sc ii} region
is estimated to be $r_{\rm S}$, 
it is useful to normalize the length scale by $r_{\rm S}$.
We assume spherical symmetry and spatial uniformity
of H {\sc ii} regions in this paper for the simplicity.

First, we calculate $\tau_{\rm S,\, d}/\tau_{\rm S,\, H}$ for
an H {\sc ii} region.
Here, both $\tau_{\rm S,\, d}$ and $\tau_{\rm S,\, H}$ are
approximated with the optical thickness at the Lyman limit
(912 \AA). 
Then we obtain
\begin{eqnarray}
\frac{\tau_{\rm S,\, d}}{\tau_{\rm S,\, H}}=
\frac{\tau_{\rm S,\, d}}{n_{\rm H}sr_{\rm S}}=12\,
\frac{E_{B-V}}{N_{\rm H}s}\, ,\label{eq:tau_ratio1}
\end{eqnarray}
where the dust extinction at 912 \AA, $A_{912}$, is taken to be
$13E_{B-V}$ mag according to the Galactic extinction curve,
and ${\tau_{\rm S,\, d}} = A_{912}/(2.5\log_{10}e)$. 
Moreover,
$n_{\rm H}$ and $s=6.30\times 10^{-18}~{\rm cm}^2$ denote the
number density of the hydrogen and the absorption cross
section for a hydrogen atom in the $n=1$ level (Eq.\ (5.6) of S78),
respectively, and the column density $N_{\rm H}$ is defined as
$N_{\rm H}\equiv n_{\rm H}r_{\rm S}$.

If the physical properties of grains such as the extinction
curve are unchanged, 
${E_{B-V}}/{N_{\rm H}}$ is proportional to the dust-to-gas mass
ratio ${\cal D}$. According to S78,
${\cal D}=6\times 10^{-3}$ when
$N_{\rm H}/E_{B-V}=5.9\times 10^{21}~{\rm mag}^{-1}~{\rm cm}^{-2}$
(Sects.\  7.2 and 7.3 of S78). Here, we apply these values to
H {\sc ii} regions. That is, we assume that  the dust-to-gas ratio in the
H {\sc ii} region is the same as that in the mean
value in the interstellar space. Then, Eq.\  (\ref{eq:tau_ratio1})
reduces to
\begin{eqnarray}
\frac{\tau_{\rm S,\, d}}{\tau_{\rm S,\, H}}=\frac{1}{3100}\,
\left(\frac{{\cal D}}{6\times 10^{-3}}\right)\, .\label{tau_ratio2}
\end{eqnarray}
This is the same as Eq.\ (5.23) in S78 but an explicit
expression of the dependence on ${\cal D}$.

Next, we calculate $\tau_{\rm S,\, H}=n_{\rm H}sr_{\rm S}$.
The Str\"{o}mgren radius is determined as
\begin{eqnarray}
r_{\rm S}=1.4\left(\frac{N_{\rm u}}{10^{48}~{\rm s}^{-1}}\right)^{1/3}
\left(\frac{n_{\rm H}}{10^2~{\rm cm}^{-3}}\right)^{-2/3}~{\rm pc}\, ,
\label{eq:stromgren}
\end{eqnarray}
where $N_{\rm u}$ represents the number of ionizing photons
emitted from central stars per second. In this equation, we have
assumed that
the temperature of the H {\sc ii} region is 8000 K
(i.e., the recombination coefficient to the $n=2$ level is
$\alpha^{(2)}=3.09\times 10^{-13}~{\rm cm}^3~{\rm s}^{-1}$;
see Eq.\ (5.14) and Table 5.2 of S78)
and that $n_{\rm e}=n_{\rm H}$, where $n_{\rm e}$ is
the number density of electrons, is satisfied  within the
Str\"{o}mgren radius.
Using the estimation of the Str\"{o}mgren radius above,
we obtain
\begin{eqnarray}
\tau_{\rm S,\, H}=2.7\times 10^3
\left(\frac{n_{\rm H}}{10^2~{\rm cm}^{-3}}\right)^{1/3}
\left(\frac{N_{\rm u}}{10^{48}~{\rm s}^{-1}}\right)^{1/3}\, ,
\end{eqnarray}
Combining this with Eq.\  (\ref{tau_ratio2}), $\tau_{\rm S,\, d}$
is estimated as
\begin{eqnarray}
\tau_{\rm S,\, d} & = & 0.87
\left(\frac{{\cal D}}{6\times 10^{-3}}\right)
\left(\frac{n_{\rm H}}{10^2~{\rm cm}^{-3}}\right)^{1/3}\nonumber \\
& & \times\left(\frac{N_{\rm u}}{10^{48}~{\rm s}^{-1}}\right)^{1/3}\, .
\label{tau_sd1}
\end{eqnarray}
We note that $\tau_{\rm S,\, d}$ becomes large as $N_{\rm u}$
increases. This is because
the probability that the dust grains absorb the photons inside an
H {\sc ii} region increases as the size of the region becomes
larger. 

\begin{figure}
 \begin{center}    
      \includegraphics[width=8cm]{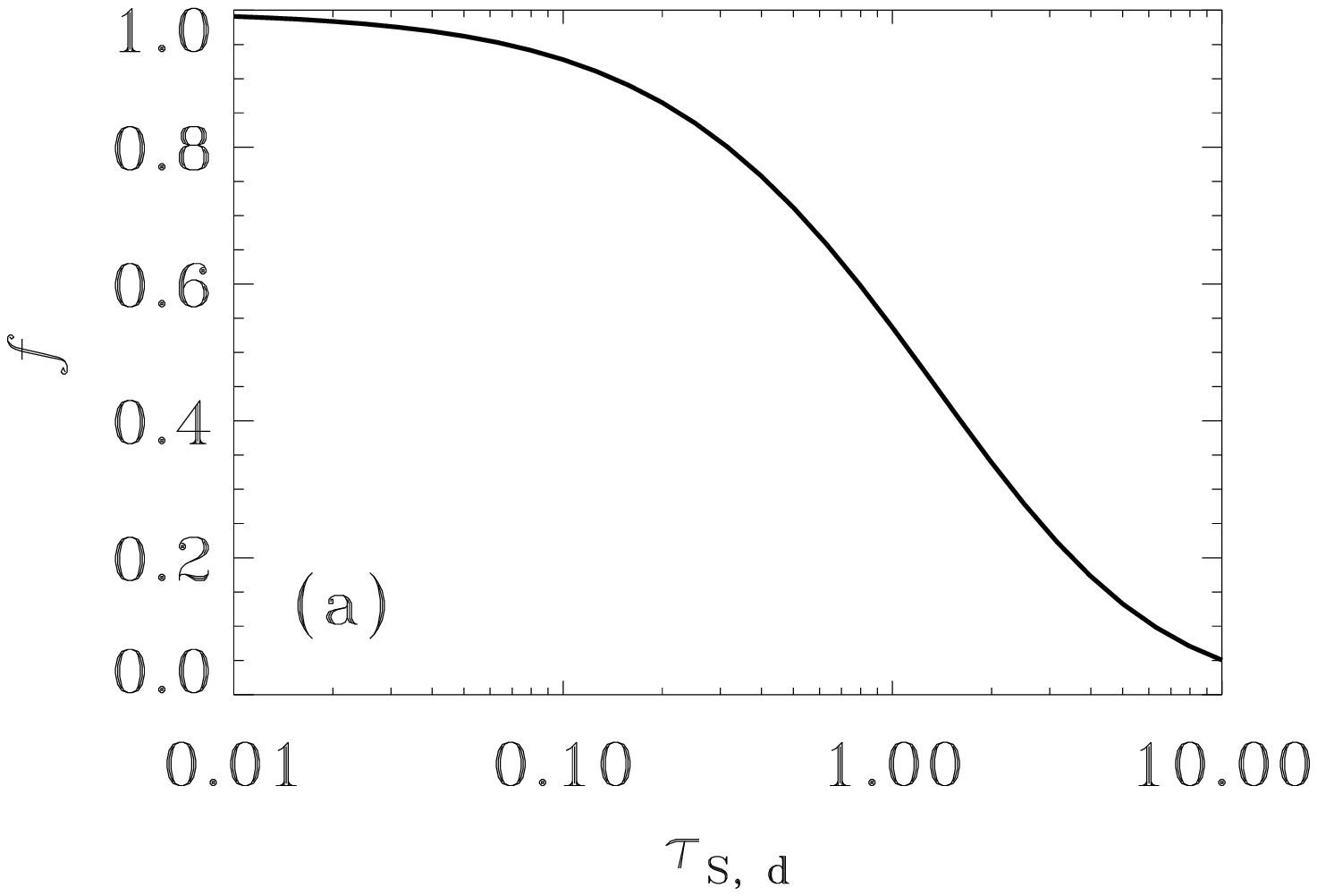}
      \includegraphics[width=8cm]{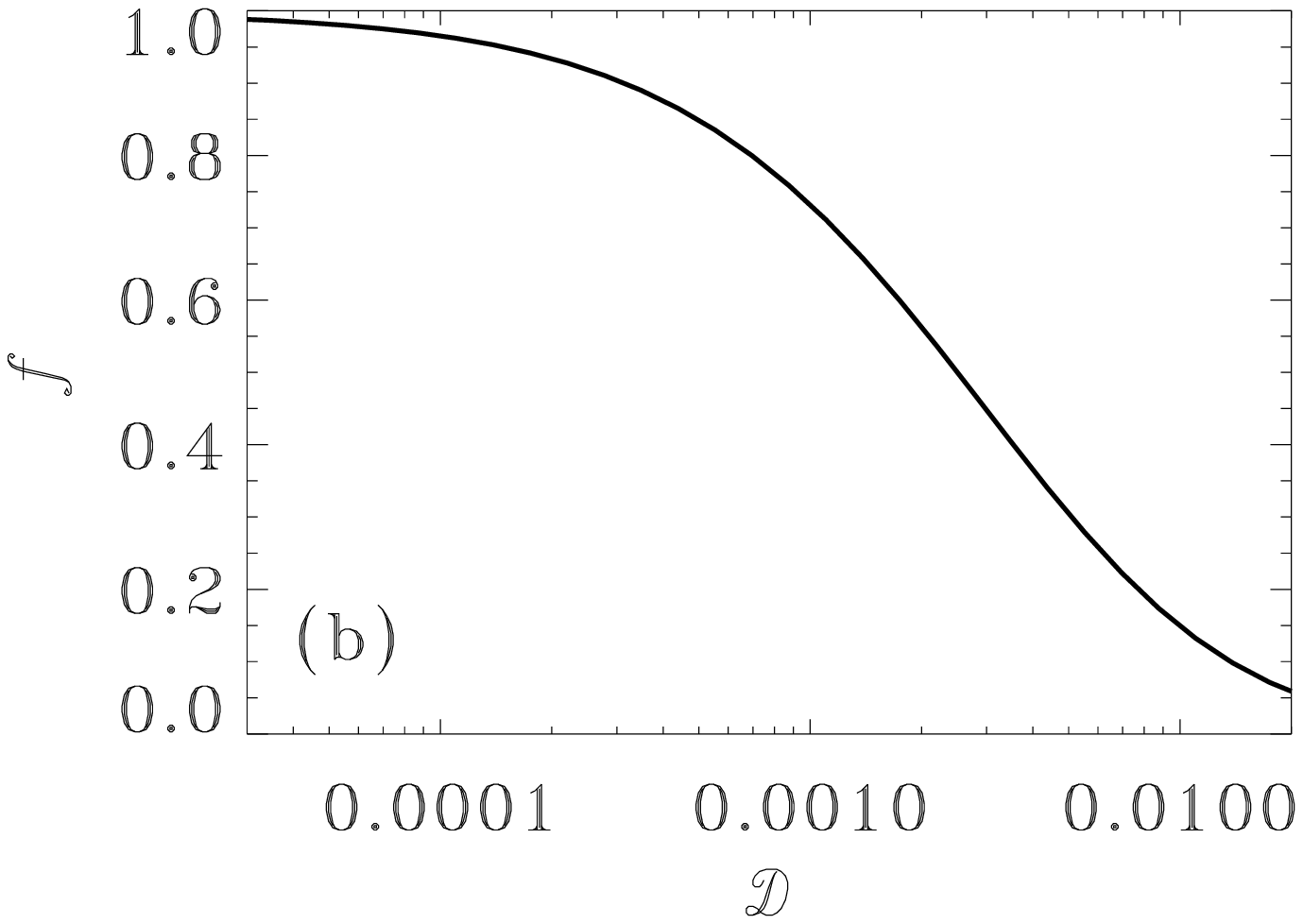}
\caption{{\bf a} Fraction of the ionizing photons
absorbed by hydrogen, $f$, as a function of $\tau_{\rm S,\, d}$
(optical depth of dust over a distance equal to the Str\"{o}mgren
radius). {\bf b} Fraction of the ionizing photons absorbed
by hydrogen, $f$, as a function of the dust-to-gas ratio,
${\cal D}$.}\label{fig1}
\end{center}
\end{figure}

Next, we estimate the fraction of the ionizing photons
absorbed by dust grains. Due to the grain absorption,
the size of an H {\sc ii} region is smaller than
$r_{\rm S}$. In other words, if we define $y_{\rm i}$ as the
ionization radius normalized by $r_{\rm S}$, $y_{\rm i}<1$.
(Without dust grains, $y_{\rm i}=1$.) A
useful relation between $\tau_{\rm S,\, d}$ and
$y_{\rm i}$ is given in Table 5.4 of S78, where
$y_{\rm i}$ is determined from the following expression
(Eq.\ (5.29) of S78):
\begin{eqnarray}
3\int_0^{y_{\rm i}}y^2e^{y\tau_{\rm S,\, d}}\,{\rm d}y=1\, .
\label{eq:implicit}
\end{eqnarray}
Using $y_{\rm i}$, the fraction of the ionizing photons
absorbed by hydrogen, $f$, is expressed as
\begin{eqnarray}
f=y_{\rm i}^3\, .
\end{eqnarray}
This is the same as $f$ in Eq.\ (\ref{inoue}).
In Fig.\  \ref{fig1}{a}, we show $f$ as a function of
$\tau_{\rm S,\, d}$.


When we consider the dependence of $f$ on the dust-to-gas ratio,
a large ambiguity exists: The number of ionizing photons per
H {\sc ii} region is unknown, since the typical number
and mass function of OB stars in an H {\sc ii} region is
difficult to determine exactly. 
Fortunately, for the purpose of finding the dependence on ${\cal D}$,
this is resolved by calibrating the ``Galactic'' $f$ with the value
of Orion Nebula (Petrosian et al.\ 1972). According to them,
$f=0.26$ in the Nebula. They also commented that the value explains
the IR emission from
H {\sc ii} regions. Adopting $f=0.26$ as the
typical value of the Galaxy (see also Inoue et al.\ \cite{inoue00b}),
we obtain 
$\tau_{\rm S,\, d}=2.7$ for the typical Galactic H {\sc ii}
regions (we consider ${\cal D}\simeq 6\times 10^{-3}$ for such regions) 
from Fig.\ \ref{fig1}{a}. Hence we write
\begin{eqnarray}
\tau_{\rm S,\, d}=2.7
\left(\frac{{\cal D}}{6\times 10^{-3}}\right)\, .
\label{eq:tau_sd2}  
\end{eqnarray}
This is consistent with Eq.\ (\ref{tau_sd1}) if we assume
$n_{\rm H} = 10^2~{\rm cm}^{-3}$ and
$N_{\rm u} \simeq 3.0\times 10^{49}~{\rm s}^{-1}$. 
Thus, the net effects of mass function and number of OB stars
are included in the value of $N_{\rm u}$ by adopting
Eq.\ (\ref{eq:tau_sd2}). 
This simplicity is meaningful for our motivation to find the
dependence of $C_{\rm IR}$ on ${\cal D}$.
Combining Eq.\  (\ref{eq:tau_sd2}) with Eq.\  (\ref{eq:implicit}),
we obtain $f$ as a function of ${\cal D}$ as shown in
Fig.\  \ref{fig1}{b}.

\subsection{Dependence of $\epsilon$ on dust-to-gas ratio}
\label{subsec:eps}

In IHK00, $\epsilon$ in Eq.\ (\ref{eq:c_ir}) is defined by
\begin{eqnarray}
\epsilon\equiv 1-e^{-\tau_{\rm nonion}}\, ,\label{eps}
\end{eqnarray}
where
$\tau_{\rm nonion}$ is the mean optical depth of dust for nonionizing
photons. That is, $\epsilon$ represents the efficiency of the
dust absorption of nonionizing photons. Inoue et al.\ (\cite{inoue00c})
estimated $\epsilon =0.9$ ($\tau_{\rm nonion} =2.3$) from
the averaged visual extinction of Usui et al.\ (\cite{usui98})'s sample 
($A_V \simeq 1$ mag)
and the Galactic extinction curve between 1000 \AA\ and 4000 \AA\ by
Savage \& Mathis (\cite{savage79}). This wavelength range is fit for
our purpose, since most of the
nonionizing photons from OB stars are emitted in the wavelength
range much
shorter than 4000 \AA. The efficient extinction in this short
wavelength makes $\epsilon$ large (i.e., nearly unity).

\begin{figure}
 \begin{center}    
      \includegraphics[width=8cm]{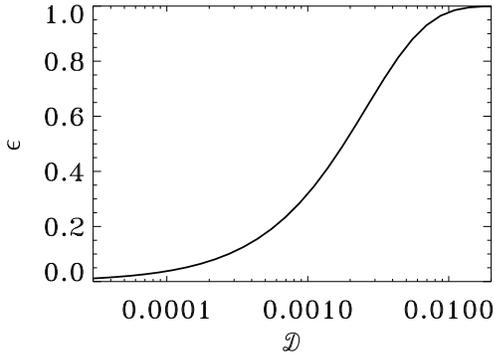}
\caption{Fraction of the nonionizing photons
absorbed by dust, $\epsilon$, as a function of the dust-to-gas
ratio, ${\cal D}$.}\label{fig2}
\end{center}
\end{figure}

It is obvious that $\tau_{\rm nonion}$ depends on the dust-to-gas
ratio. If the column density of gas contributing to the absorption
of nonionizing photons is fixed, $\tau_{\rm nonion}$ is proportional
to the dust-to-gas ratio. Since we are interested in the dependence of
$C_{\rm IR}$ on ${\cal D}$, we simply adopt
$\tau_{\rm nonion}\propto{\cal D}$. Here, we determine the numerical
value of $\tau_{\rm nonion}$ as
\begin{eqnarray}
\tau_{\rm nonion}=2.3\left(\frac{{\cal D}}{6\times 10^{-3}}\right)\, ,
\label{tau_nonion}
\end{eqnarray}
so that $\tau_{\rm nonion}$ becomes 2.3 for the Galactic
dust-to-gas ratio.
By combining Eqs.\ (\ref{eps}) and (\ref{tau_nonion}),
we obtain $\epsilon$ as a function of
${\cal D}$ as shown in Fig.\  \ref{fig2}.

\subsection{Cirrus Fraction}\label{subsec:cirrus}

The fraction of the cirrus component, $\eta$, remains to be
determined. In this paper, 
an empirical value of $\eta$ is simply adopted.
According to Lonsdale Persson \& Helou (\cite{lonsdale87}),
$\eta\sim 0.5$--0.7 for their
sample spiral galaxies. We use $\eta =0.5$ as adopted in IHK00
(the value averaged for Usui et al.\ 1998's sample) for the current
estimate for our SFR or $C_{\rm IR}$.
As a first step, we adopt a constant $\eta$, since
a theoretically reasonable determination of it is very difficult.
The variation of the cirrus fraction by the change of
${\cal D}$ will be briefly considered in Appendix A.
We note that the range of $C_{\rm IR}$ becomes larger
if we consider the metallicity dependence of the cirrus fraction
(Appendix A).

\subsection{Absorption of Lyman-$\alpha$ photons by dust}

Since IHK00's derivation of Eq.\  (\ref{inoue}) is based
on Case B, Ly$\alpha$ photons are assumed to be easily trapped in
an H {\sc ii} region (Osterbrock \cite{osterbrock89}).
Thus, during the resonant
scatterings in an H {\sc ii} region, all the Ly$\alpha$
photons are assumed to be absorbed by grains in IHK00, which
this paper is based on.

If the dust-to-gas ratio is significantly smaller than 
the Galactic value, the dust grains might not efficiently absorb
the Ly$\alpha$ photons in H {\sc ii} regions.
However, H {\sc i} envelopes on a galactic scale generally
exists around
H {\sc ii} regions. A Ly$\alpha$ photon is
absorbed in the H {\sc i} envelopes soon after
it escapes from an H {\sc ii} region (Osterbrock \cite{osterbrock61}).
Hence, even if there is only a small amount of dust,
we expect a sufficient chance for dust to absorb the Ly$\alpha$
photons in the H {\sc i} envelope.
We check this point below.

First of all, we define a path length of the Ly$\alpha$ photons
as $l_{\rm path}$. It is estimated as being 
$l_{\rm path}\sim\tau_{{\rm Ly}\alpha}^2l_{{\rm Ly}\alpha}$,
where $\tau_{{\rm Ly}\alpha}$ means the optical depth for
Ly$\alpha$ photons and $l_{{\rm Ly}\alpha}$ is the mean free path of the
Ly$\alpha$ photons. 
The square to $\tau_{{\rm Ly}\alpha}$ means that the resonant scattering
is assumed to be a random-walk process of photons. Next, we define an
optical depth of dust grains for Ly$\alpha$ photons  over the length of
$l_{\rm path}$ as
$\tau_{\rm dust}$. It is estimated as
$\tau_{\rm dust} \sim \pi a^2 l_{\rm path} n_{\rm dust}$, where
$a$ is the size of grains and $n_{\rm dust}$ is their number density.
Here, we will discuss whether the Ly$\alpha$ photons have a chance to
escape from galaxies. We are interested in a star-forming region
surrounded by H {\sc i} gas envelopes.
The scale length of the H {\sc i} envelope, $L$, may be
estimated to be about 100 pc, 
which corresponds to the thickness of the disk of spiral galaxies.
In such a case, $\tau_{{\rm Ly}\alpha}$ is estimated to be about
$L/l_{{\rm Ly }\alpha}$. Then, we find
$\tau_{\rm dust} \sim \pi a^2 n_{\rm dust} L^2 /l_{{\rm Ly }\alpha}$.
Using ${\cal D}$, it is expressed as
\begin{eqnarray}
\tau_{\rm dust} \sim 
\frac{\pi a^2 n_{\rm H} {\cal D} L^2 m_{\rm p} }
{ l_{{\rm Ly }\alpha} m_{\rm dust} }\, ,
\end{eqnarray}
where $m_{\rm p}$ is the proton mass and $m_{\rm dust}$ is the dust
mass for an assumed spherical dust with mass density of 3 g cm$^{-3}$.
Adopting $n_{\rm H} = 0.1$ cm$^{-3}$ for a diffuse 
H {\sc i} medium, $a=0.1$ $\mu$m,
${\cal D} = 6 \times 10^{-5}$ (0.01 times smaller than the Galactic
value), $L= 100$ pc,
and the resonant-scattering cross section of Ly$\alpha$ photons
$2 \times 10^{-13}$ cm$^{2}$ (i.e.,
$l_{{\rm Ly }\alpha}=5\times 10^{13}$ cm),
we find $\tau_{\rm dust} \sim 5\times 10^2$. Thus, $\tau_{\rm dust}$
is much larger than unity.
This means that most of the Ly$\alpha$ photons are absorbed
even if the dust-to-gas ratio is as small as 0.01 times the
Galactic value. 

\subsection{Dependence of $C_{\rm IR}$ on dust-to-gas ratio}

\begin{figure}
 \begin{center}    
      \includegraphics[width=8cm]{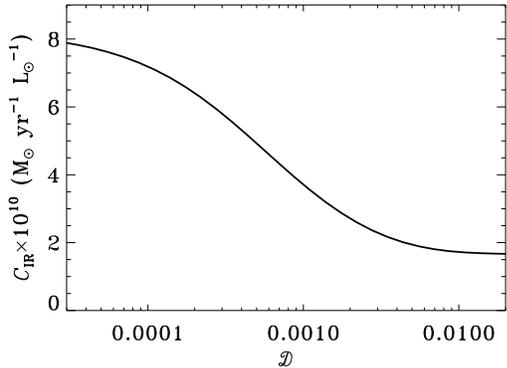}
\caption{$C_{\rm IR}$ (conversion coefficient
from the IR luminosity of dust to the star formation rate) as a 
function of the dust-to-gas ratio, ${\cal D}$.}\label{fig3}
\end{center}
\end{figure}

In the above subsections, we have expressed $f$ and $\epsilon$
as a function of the dust-to-gas ratio (${\cal D}$),
while $\eta$ is treated as a constant ($\eta =0.5$).
We also assume a typical star-forming region whose mean density
of gas and production rate of ionizing photons are
about $10^2$ cm$^{-3}$ and $3.0\times 10^{49}$ s$^{-1}$,
respectively. Then, we can express $C_{\rm IR}$ defined in
Eq.\ (\ref{eq:c_ir})
as a function of the dust-to-gas ratio. In Fig.\  \ref{fig3},
we present $C_{\rm IR}$ as a function of ${\cal D}$.
{}From this figure, we find that
the coefficient of the
conversion from IR light to SFR becomes 4--5 times smaller
for ${\cal D}\sim 6\times 10^{-3}$ (the Galactic value) than that
for ${\cal D}\sim 6\times 10^{-5}$ (0.01 times the Galactic value).

$C_{\rm IR}$ is also changed if we adopt a different extinction curve.
In the previous discussions,
we have adopted the extinction curve of the Galaxy.
The current paragraph examines how the relation between
$C_{\rm IR}$ and ${\cal D}$ changes when we adopt the
extinction curve
of the Small Magellanic Cloud (SMC) with the other quantities adopted
in this paper unchanged.
In the SMC-type extinction law,
the ratio of the optical depth in Eq.\ (\ref{eq:tau_ratio1}) becomes
two times
larger, because the UV extinction is enhanced in the SMC extinction.
This indicates that
the same estimation performed in Sect.\ \ref{sec:sfr} is possible if
we make the normalization
of ${\cal D}$ half. Thus, we can find a rough dependence of $f$ on
${\cal D}$ from Fig.\ \ref{fig1}b by doubling ${\cal D}$
(i.e., the line in Fig.\ \ref{fig1}b is moved to the left
by 0.3 dex). 
Since ${\cal D}$ of SMC is about $5\times 10^{-4}$
(Issa et al.\ \cite{issa90}), 
we find $f\simeq 0.75$ for the SMC extinction case, while
$f\simeq 0.85$
if we adopt the extinction law of the Galaxy.
We can apply the same scaling to $\epsilon$.
For ${\cal D}\simeq 5\times 10^{-4}$, $\epsilon\simeq 0.4$ for
the SMC extinction case, while $\epsilon\simeq 0.2$ for the
Galactic extinction case. As a result, at ${\cal D}\simeq 5\times 10^{-4}$,
$C_{\rm IR}\simeq 3.4\times 10^{-10}~M_\odot~{\rm yr}^{-1}~L_\odot^{-1}$
for the SMC extinction, while 
$C_{\rm IR}\simeq 4.7\times 10^{-10}~M_\odot~{\rm yr}^{-1}~L_\odot^{-1}$
for the Galactic extinction.
Thus, we should be aware of the effect of extinction law
on our conversion formula, but the effect is small (a factor of
1.4).
If we remember that the uncertainty owing to the IMF is a factor of 2
(e.g., IHK00), 
we find that the dependence of $C_{\rm IR}$ on the 
the extinction curve is not very important.
We note that we can estimate $C_{\rm IR}$ for the SMC-type extinction
by moving the line in Fig.\ \ref{fig3} to the left by 0.3 dex
(a factor of 2).

\subsection{Observational check}\label{subsec:check}

Here, we examine whether the values of $f$ and $\epsilon$
that we adopted for the Galactic value (i.e., $f=0.26$ and
$\epsilon =0.9$) are consistent with the properties of actual
H {\sc ii} regions, by using the ratio between
the ionizing-photon luminosity and the IR luminosity.
This ratio was extensively studied in 1970s (e.g.,
Harper \& Low \cite{harper71}).

We start with the relation by Petrosian et al.\ (\cite{petrosian72})
as
\begin{eqnarray}
L_{\rm IR}=L({\rm Ly}\alpha)+(1-f)\langle h\nu\rangle_{\rm ion}N_{\rm u}
+\epsilon L_{\rm nonion}\, ,\label{eq:petrosian}
\end{eqnarray}
where $\langle h\nu\rangle_{\rm ion}$ is the mean energy of 
an ionizing photon, and $L_{\rm IR}$, $L({\rm Ly}\alpha)$, and
$L_{\rm nonion}$ are luminosities of IR dust emission,
Ly$\alpha$, and nonionizing photons, respectively. We note that
the meanings of $f$ and $\epsilon$ in Eq.\ (\ref{eq:petrosian})
are the same as those in Eq. (\ref{eq:c_ir}).
IHK00 derived $L_{\rm nonion}=1.5\langle h\nu\rangle_{\rm ion}N_{\rm u}$
by adopting the Salpeter's initial mass function
(Salpeter \cite{salpeter55}) with the stellar mass range of
0.1--100 $M_\odot$ and the mass--luminosity relation of stars
(Table 3.13 of Binney \& Merrifield \cite{binney98}).
Then, according to Mezger (\cite{mezger78}), we define the IR
excess (IRE) as
\begin{eqnarray}
{\rm IRE}\equiv\frac{L_{\rm IR}}{h\nu_{{\rm Ly}\alpha}fN_{\rm u}}=a+(1-f
+1.5\epsilon )\,
\frac{\langle h\nu\rangle_{\rm ion}}{h\nu_{{\rm Ly}\alpha}f}\, ,
\end{eqnarray}
where $\nu_{{\rm Ly}\alpha}$ is the frequency of the Ly$\alpha$
radiation, and $a$ is the fraction of Ly$\alpha$ photons that reach the 2p
state and go down to 1s (i.e.,
$L({\rm Ly}\alpha)=
ah\nu_{{\rm Ly}\alpha}fN_{\rm u}$)\footnote{The other photons decay to
the ground state with a two-photon process.}. According to S78,
$a\simeq 0.67$. Adopting $f=0.26$ and $\epsilon =0.9$ for the
Galactic values (Sects.\ \ref{subsec:f} and \ref{subsec:eps}),
we obtain ${\rm IRE}=11$, where we have assumed that
$\langle h\nu\rangle_{\rm ion}$ is equal to $h\nu$ at the
Lyman limit (912 \AA). According to Harper \& Low (\cite{harper71}),
$5\la{\rm IRE}\la 10$ (see also Aannestad \cite{aannestad78};
Mezger \cite{mezger78}; Maihara et al.\ \cite{maihara81}) for
Galactic H {\sc ii} regions. Fig.\ 7.3 in S78 showed that
the IR luminosity of H {\sc ii} regions is larger than predicted
from Ly$\alpha$ luminosity by an order of magnitude (i.e.,
${\rm IRE}\sim 10$).
Thus, $f$ and $\epsilon$ adopted in
this paper are almost consistent with the properties of 
the Galactic H {\sc ii} regions within the scatter of 
observed IRE.

Another test of our model is possible for the case of H {\sc ii}
regions in the Large Magellanic Cloud
(LMC) by using the observational results of
DeGioia-Eastwood (\cite{degioia92}).
{}From Table 3 of DeGioia-Eastwood (1992), we obtain
systematically lower IRE for the LMC H {\sc ii}
regions; IRE ranges from 1 to 4, and the median is 1.6.
This can be naturally explained by our model with a lower value
of the dust-to-gas ratio in the LMC. If we adopt
${\cal D}=1.2\times 10^{-3}$ for the LMC (one-fifth of that
of the galaxy) according to Issa et al.\ (1990), we obtain
$f\simeq 0.7$ and $\epsilon\simeq 0.4$.
Then we obtain ${\rm IRE}\simeq 2.4$ for
the LMC. 
This value of IRE for the LMC lies in the range
of the observed IRE.
Thus, we consider Eqs.\ (\ref{eq:tau_sd2}) and (\ref{tau_nonion}) to be
applicable and useful in spite of simplification in their derivation.
Therefore, we discuss the metallicity dependence of
$C_{\rm IR}$ and its 
effect on the cosmic star formation history by using the relation 
between $C_{\rm IR}$ and ${\cal D}$ as shown in Fig.\
\ref{fig3} in the following sections.

\section{Metallicity dependence}\label{sec:metal}

\begin{figure}
 \begin{center}    
      \includegraphics[width=8cm]{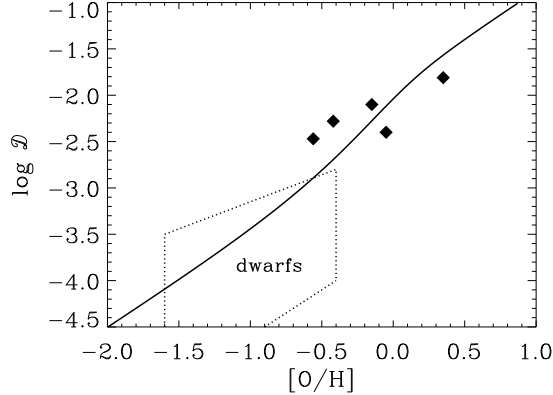}
\caption{Relation between the dust-to-gas ratio
${\cal D}$ and oxygen abundance [O/H]. The solid line represents the
well-fitting model
by Hirashita (1999a, b). The observed relations
for nearby spiral galaxies are presented with the square
(Issa et al.\ 1990).
The area marked with dwarfs represents a typical locus of
dwarf irregular galaxies and blue compact dwarf galaxies
(Lisenfeld \& Ferrara \cite{lisenfeld98}).}\label{fig4}
\end{center}
\end{figure}

In order to obtain the metallicity dependence of the conversion
formula (Eq.\ (\ref{inoue})), which depends on the dust-to-gas ratio,
${\cal D}$, as
shown in the previous section, we must relate ${\cal D}$ and the
metallicity. Indeed, ${\cal D}$ and metallicity are positively
correlated (Issa et al.\ \cite{issa90};
Schmidt \& Boller \cite{schmidt93};
Lisenfeld \& Ferrara \cite{lisenfeld98}).
Here, we adopt the latest relation proposed by
Hirashita (1999a, b).
He has constructed a new evolution model of dust in the galactic
environment, emphasizing the importance that
the grains can grow via the accretion of the metal elements 
in the cool and neutral components of interstellar medium (ISM;
see also Dwek \cite{dwek98}).
Then, his relation between ${\cal D}$ and metallicity has explained
the observational relation for both the giant and dwarf galaxies.
Here, we adopt the relation as shown in Fig.\ \ref{fig4}
(solid line), since
this seems to reproduce the relation for dwarf galaxies and
spiral galaxies\footnote{As for the parameters in Hirashita (1999a, b), we
choose the values as
$f_{\rm in,\, O}=0.1$ and $\beta_{\rm acc}=2\beta_{\rm SN}=10$, where
$f_{\rm in,\, O}$ is the dust mass fraction in the material injected
from stars, and $\beta_{\rm acc}$ and $\beta_{\rm SN}$ are
defined as gas consumption timescale (gas mass divided by star
formation rate) normalized by dust growth timescale in clouds and by
dust destruction timescale in supernova shocks, respectively (see
Hirashita 1999a for details).}. The metallicity is represented by
[O/H]. We note that
${\rm [O/H]}=x$ means that the abundance of oxygen is $10^x$ times
the solar value (the solar oxygen abundance is assumed to be
0.013 in mass; e.g., Whittet \cite{whittet92}).
In Fig.\ \ref{fig4}, observational data are also
shown: The observed relations
for nearby spiral galaxies are presented with the square
(Issa et al.\ \cite{issa90}), and
the area marked with dwarfs represents a typical locus of
dwarf irregular galaxies and blue compact dwarf galaxies
(Lisenfeld \& Ferrara \cite{lisenfeld98}). We hereafter
adopt the solid line in Fig. \ref{fig4} as the relation between the
dust-to-gas ratio and the metallicity.

\begin{figure}
 \begin{center}    
      \includegraphics[width=8cm]{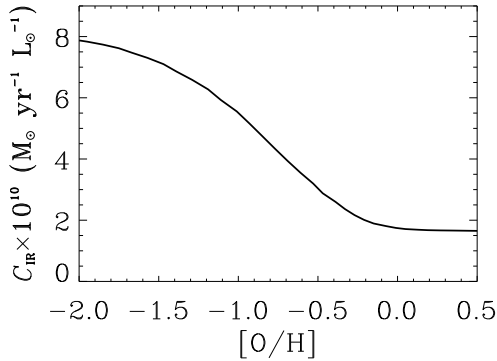}
\caption{$C_{\rm IR}$ (conversion coefficient
from the IR luminosity of dust to the star formation rate) as a 
function of the oxygen abundance, [O/H].}\label{fig5}
\end{center}
\end{figure}

Once we accept the relation in Fig.\  \ref{fig4}, we relate
$C_{\rm IR}$ and [O/H] by using the relation between the dust-to-gas
ratio and $C_{\rm IR}$ (Fig.\ \ref{fig3}). The relation between
$C_{\rm IR}$ and [O/H] is presented in Fig.\  \ref{fig5}. We see
that if [O/H] in the ISM evolves
from $-2$ to 0 via chemical evolution, the coefficient of the
conversion from IR light to SFR becomes about 4--5 times smaller,
again. Thus, when we would like to determine the SFR precisely within a 
factor of 4--5, we should not neglect the effect of metallicity.

\section{Comments on the cosmic star formation history}
\label{sec:cosmic}

The cosmological evolution of SFR is derived
from the comoving density of galactic light. For example,
Madau et al.\ (\cite{madau96}) applied the conversion formula from UV
light to the SFR and showed that the SFR as a function of the
redshift $z$ seems to have a peak at $z\sim 1$--2.
The cosmic SFR as a function of $z$ has been revised and discussed
by many authors (e.g., Steidel et al.\ \cite{steidel99}), commonly
referred to as the Madau plot.
Takeuchi et al.\ (\cite{t2_00}) applied the formula by
IHK00 to the determination of the cosmic star formation history
from the number-count data of IR galaxies.

The cosmic SFR is also determined from
the dust emission at the sub-millimeter (sub-mm) observational
wavelength
(Hughes et al.\ \cite{hughes98}; Barger et al.\ \cite{barger98}).
We can convert the sub-mm luminosity to the SFR by applying the formula
proposed in IHK00. If we take into account the
chemical evolution of galaxies on a cosmological timescale
(e.g., Pei \& Fall \cite{pei95}), however, we must examine the
dependence of the conversion formula on 
metallicity as seen in the previous
sections.

According to such a motivation,
we apply the results obtained in the previous sections, especially
Fig.\  \ref{fig5} and its related discussions, to the cosmic star
formation history.
We consider the evolution of our conversion 
law along the metal enrichment with the aid of previous research
that determined the metallicity as a function of $z$. We adopt
Pei et al.\ (\cite{pei99})
as a recent modeling of the cosmic chemical evolution.

Pei et al.\ (\cite{pei99}) modeled the cosmic star formation
history and chemical evolution and calculated the evolution of
dust amount. Considering the absorption and reprocessing of
light by dust, they determined the cosmic chemical evolution
after the calibration with the observed comoving density of
light. The resultant metallicity
evolution as a 
function of $z$ is shown in the second column of Table \ref{tab1}.
Here we adopt the solid line of Fig.\ 8 in Pei et al.\ (\cite{pei99}).
Though their treatment of the evolution of the dust-to-gas
ratio is not just the same as our treatment, 
for readers' qualitative understanding, we present a relation 
between the metallicity and the redshift $z$ from their Fig.\  8.


\begin{table}
 \caption[]{Metallicity and $C_{\rm IR}$ as functions of
     $z$.}\label{tab1}
 \begin{tabular}{ccc}
  \hline
   $z$ & $\log (Z/Z_\odot )$ & $C_{\rm IR}$ \\
       &                     & [$M_\odot~{\rm yr}^{-1}~L_\odot^{-1}$]
   \\ \hline
   0.0 & $~~0.00$ & $1.7\times 10^{-10}$ \\
   0.5 & $-0.22$  & $2.0\times 10^{-10}$ \\
   1.0 & $-0.48$  & $2.9\times 10^{-10}$ \\
   2.0 & $-1.02$ & $5.6\times 10^{-10}$ \\
   3.0 & $-1.50$ & $7.1\times 10^{-10}$ \\
   4.0 & $-1.79$ & $7.6\times 10^{-10}$ \\
   5.0 & $-1.88$ & $7.8\times 10^{-10}$ \\
 \hline
 \end{tabular}
\end{table}

We also present $C_{\rm IR}$ in the third column of Table \ref{tab1} 
calculated from
the relation between metallicity ($Z$) and $C_{\rm IR}$ as shown in
Fig.\  \ref{fig5}. Here we assume that
$\log (Z/Z_\odot)=[{\rm O}/{\rm H}]$, where $Z_\odot$ is the solar
metallicity and $Z$ is measured in the galactic scale.
{}From Table \ref{tab1}, we see that $C_{\rm IR}$
is 4--5 times larger at $z\sim 3$ than that at
$z\sim 0$. Thus, we should carefully consider the chemical evolution
of the galaxies if we determine the cosmic star formation
history from the dust emission within a factor of 4--5.
Deriving the cosmic star formation history from IR--sub-mm
observation will be made possible by the future
observational projects (e.g., Takeuchi et al.\ \cite{t2_99}).
Our formulation, which takes into account the metallicity dependence,
will be useful in determining
the cosmic star formation history.

We should mention that Pei et al.\ (\cite{pei99}) treated
averaged quantities for each redshift and focused only on the redshift
dependence. Thus, our Table \ref{tab1} must be applied to the
data averaged
for each $z$. In other words, we should not apply them
to each individual galaxy at a certain $z$. For each galaxy,
Fig.\ \ref{fig5} should be used instead after their metallicity
is known. Finally, we should keep in mind the possibility that
far-IR (FIR) or sub-mm sample of galaxies is biased to metal-enriched
system because metal-poor (i.e., dust-poor) galaxies are not easily
observed in these wavelengths.

\section{Conclusion}\label{sec:conclusion}

Based on IHK00's formulation, we consider the factor $C_{\rm IR}$
in the conversion formula between the IR luminosity and the
SFR (Eqs.\ (\ref{eq:conversion}) and
(\ref{eq:c_ir})). The factor $C_{\rm IR}$ becomes
$1.7\times 10^{-10}$, $5.6\times 10^{-10}$, and
$7.9\times 10^{-10}$ $M_\odot~{\rm yr}^{-1}~L_\odot^{-1}$
for the metallicity of 1, 0.1, 0.01 times the solar value,
respectively. Thus, $C_{\rm IR}$ differs by a factor of 4--5 in
the range. 
Importantly, applicability of our formula is observationally
confirmed by IREs of the Galaxy and the LMC.
Applying our result to the cosmic star formation
history, we have found that $C_{\rm IR}$
may be about 4--5 times larger at $z\sim 3$ than that at
$z\sim 0$. This factor of 4--5 is larger than the uncertainty caused
by the initial mass function of stars (a factor of 2).
Thus, we should carefully consider the chemical
evolution of galaxies if we determine the cosmic star formation
history from the dust emission within a factor of 4--5.

\begin{acknowledgements}
We first thank U. Lisenfeld, the referee, for useful comment that
much improved the quality of this paper.
We acknowledge T. T. Takeuchi and T. Totani for
helpful discussions and S. Mineshige for continuous
encouragement. One of us (HH) acknowledges the Research Fellowship of
the Japan
Society for the Promotion of Science for Young Scientists.
We made extensive use of
the NASA's Astrophysics Data System Abstract Service (ADS).
\end{acknowledgements}

\appendix

\section{dependence of cirrus fraction on metallicity}

In the main body of this paper, 
we have assumed that the cirrus fraction of the IR
luminosity, $\eta$, is 0.5. 
Here, we examine how $C_{\rm IR}$ is affected owing to
the change of $\eta$.
There is a large uncertainty about observational
estimate of  $\eta$. But, theoretically, we consider some simple
cases 
where the change in $\eta$ cannot be neglected in our conversion
formula.

\begin{figure}
 \begin{center}    
      \includegraphics[width=8cm]{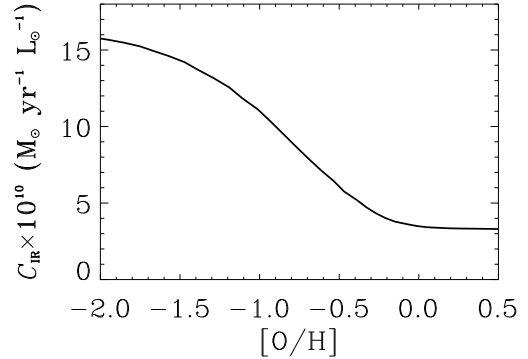}
\caption{$C_{\rm IR}$ (conversion coefficient
from the IR dust luminosity to the star formation rate) as a 
function of the oxygen abundance, [O/H], for starburst galaxies. The
cirrus fraction $\eta$ is assumed to be 0.}\label{figa1}
\end{center}
\end{figure}

The assumption of $\eta =0.5$ breaks if we are interested in
starburst galaxies.
When we examine a sample of starburst galaxies, indeed, 
it is reasonable to assume $\eta\simeq 0$. This is because
young stars dominate the radiation field that heats the
dust, and the optical depth of dust is so large that
almost all of the bolometric luminosity is emitted in
the IR (Soifer et al.\ \cite{soifer87}; Kennicutt \cite{kennicutt98b}).
For the starburst galaxies, thus,
the dependence of $C_{\rm IR}$ on the metallicity is
obtained by putting $\eta =0$ into Eq.\ (\ref{eq:c_ir}), while the
dependence of
$f$ and $\epsilon$ on the metallicity is the same as described in
Sect.\ \ref{sec:metal}. The result is shown in Fig.\ \ref{figa1}.
We see that $C_{\rm IR}$ in this figure is larger by a factor of 2
than that in Fig.\ \ref{fig5}.

\begin{figure}
 \begin{center}    
      \includegraphics[width=8cm]{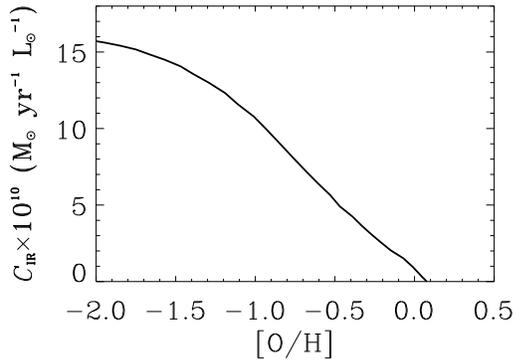}
\caption{$C_{\rm IR}$ (conversion coefficient
from the IR dust luminosity to the star formation rate) as a 
function of the oxygen abundance, [O/H], for spiral galaxies. The
cirrus fraction $\eta$ is assumed to be proportional to the
dust-to-gas ratio.}\label{figa2}
\end{center}
\end{figure}

Next, we consider a possible change in $\eta$ of spiral galaxies
as a function of
metallicity. The cirrus fraction of spiral galaxies is
not negligible at the present epoch ($\eta\simeq 0.5$
for nearby spiral galaxies;
Lonsdale Persson \& Helou \cite{lonsdale87}).
In the beginning of a spiral galaxy's evolution, when only a little
metal is produced,
$\eta=0.0$ may be reasonable.
Thus, the
first approximation for the dependence of $\eta$ on the dust-to-gas
ratio is
\begin{eqnarray}
\eta =0.5\left(\frac{\cal D}{6\times 10^{-3}}\right)\, .\label{eq:cirrus}
\end{eqnarray}
If we adopt this relation, we obtain Fig.\ \ref{figa2} as the
relation between $C_{\rm IR}$ and metallicity. 
We note that our simple treatment as Eq.\ (\ref{eq:cirrus})
is only applicable for ${\rm [O/H]}\la 0.1$, because $\eta >1$
does not make sense.
This means that we need a nonlinear modeling of the relation between
$\eta $ and ${\cal D}$ at the present epoch.
We may need to consider a complex mode of evolution of ISM,
whose nonlinearity causes intermittent star formation history
(Kamaya \& Takeuchi \cite{kamaya97}; 
Hirashita \& Kamaya \cite{hirashitakamaya00};
Takeuchi \& Hirashita \cite{t2hirashita00}). 
Comparing $C_{\rm IR}$ at
${\rm [O/H]}=-2$ with that at ${\rm [O/H]}=0$, we see that
$C_{\rm IR}$ changes by an order of magnitude as the chemical
enrichment proceeds.
Thus, if we want to know the realistic cosmic evolution of galaxies
from $L_{\rm IR}$, the metallicity evolution of cirrus must also be
examined. We will try this very difficult theme in the near future.

\end{document}